\documentclass[twocolumn]{article}

\usepackage{PRIMEarxiv}

\usepackage[utf8]{inputenc} % allow utf-8 input
\usepackage[T1]{fontenc}    % use 8-bit T1 fonts
\usepackage{hyperref}       % hyperlinks
\usepackage{url}            % simple URL typesetting
\usepackage{booktabs}       % professional-quality tables
\usepackage{amsfonts}       % blackboard math symbols
\usepackage{nicefrac}       % compact symbols for 1/2, etc.
\usepackage{microtype}      % microtypography
\usepackage{lipsum}
\usepackage{fancyhdr}       % header
\usepackage{graphicx}       % graphics

\usepackage{amsmath,amssymb,amsfonts}
\usepackage{algorithmic}
\usepackage{graphicx}
\usepackage{textcomp}
\usepackage{mathptmx} 
\usepackage{multirow}
\usepackage{enumerate}
\usepackage{color,soul}
\sethlcolor{yellow}

\usepackage{times}
\usepackage{subcaption}

\usepackage[numbers]{natbib}
\usepackage{booktabs}
\usepackage{lscape}
\usepackage{comment}
\usepackage{verbatim}
\usepackage{ltablex}

\usepackage{stfloats}
\usepackage{lscape}

\usepackage{wrapfig}

\usepackage{longtable}

\usepackage{varwidth}
\usepackage{xcolor, colortbl}

\usepackage{enumitem}

\usepackage{orcidlink}

\title{Enhancing Adversarial Robustness of IoT Intrusion Detection via SHAP-Based Attribution Fingerprinting} 

\author{
 Dilli Prasad Sharma {\orcidlink{0000-0001-5497-3850}}, Liang Xue,  Xiaowei Sun\\
  School of Information Technology, York University, Toronto, Ontario, Canada \\
  \texttt{\{dilli, lxue03, xiaoweis\}@yorku.ca}  \\
  \and
  Xiaodong Lin \\
  School of Computer Science, University of Guelph, Ontario, Canada \\ 
  \texttt{xlin08@uoguelph.ca} \\
  \and
 Pulei Xiong \\
  National Research Council of Canada, Ottawa, Ontario, Canada \\
  \texttt{pulei.xiong@nrc-cnrc.gc.ca}   
}

\begin{document}
\twocolumn[ {%
\begin{@twocolumnfalse}
\maketitle
 \begin{abstract}
		The rapid proliferation of Internet of Things (IoT) devices has transformed numerous industries by enabling seamless connectivity and data-driven automation. However, this expansion has also exposed IoT networks to increasingly sophisticated security threats, including adversarial attacks targeting artificial intelligence (AI) and machine learning (ML)-based intrusion detection systems (IDS) to deliberately evade detection, induce misclassification, and systematically undermine the reliability and integrity of security defenses. To address these challenges, we propose a novel adversarial detection model that enhances the robustness of IoT IDS against adversarial attacks through SHapley Additive exPlanations (SHAP)-based fingerprinting. Using SHAP’s DeepExplainer, we extract attribution fingerprints from network traffic features, enabling the IDS to reliably distinguish between clean and adversarially perturbed inputs. By capturing subtle attribution patterns, the model becomes more resilient to evasion attempts and adversarial manipulations. We evaluated the model on a standard IoT benchmark dataset, where it significantly outperformed a state-of-the-art method in detecting adversarial attacks. In addition to enhanced robustness, this approach improves model transparency and interpretability, thereby increasing trust in the IDS through explainable AI.
	\end{abstract}    
	
\end{@twocolumnfalse} 
}
]
% ----------------------- Preprint notice as bottom-of-page footnote -----------------------
\renewcommand{\thefootnote}{} % remove footnote number
\footnotetext[0]{%

\vspace{2mm}
This work has been submitted to the IEEE for possible publication. Copyright may be transferred without notice, after which this version may no longer be accessible.
}
\renewcommand{\thefootnote}{\arabic{footnote}} % restore footnote numbering

%\footnotetext[']{*This work has been submitted to the IEEE for possible publication. Copyright may be transferred without notice, after which this version may no longer be accessible.}
	\section{Introduction}
	\label{section:intro}
	The Internet of Things (IoT) has rapidly emerged as a transformative paradigm, revolutionizing industries such as healthcare, manufacturing, transportation, and smart infrastructure by enabling real-time data collection, seamless interconnectivity, and intelligent automation~\cite{Al-Fuqaha2015IoT}. However, the pervasive deployment of IoT devices and their reliance on heterogeneous, resource-constrained environments have significantly expanded the attack surface for malicious actors such as distributed denial of service (DDoS), side-channel, false data injection, man-in-the-middle, sniffing, spoofing, and Mirai botnet  attacks~\cite{Hassija2019ASurvey, Mishra2021Internet}. In response, advanced machine learning (ML) and deep learning (DL) techniques have been widely adopted in modern intrusion detection systems (IDS) for their ability to model complex patterns and detect sophisticated attacks in network traffic~\cite{Zhang2022Adversarial, Chen2022Machine}. Despite these advancements, DL-based IDSs are often criticized for their lack of interpretability, functioning as black-box models with opaque and non-intuitive internal decision-making processes~\cite{ Guo2018Lemna}. This inherent opacity not only hinders trust and accountability but also makes them vulnerable to adversarial attacks~\cite{Chander2025Toward, Zhang2025Explainable, Szegedy2013Intriguing, Qiu2020Adversarial}. In these attacks, attackers deliberately manipulate input data to induce misclassifications or evade detection, which in turn critically undermines the reliability and robustness of the IDSs~\cite{Alhajjar2021Adversarial, He2023Adversarial}.

    To address adversarial vulnerabilities in deep learning-based intrusion detection systems, researchers have proposed a wide range of adversarial defense methods, including adversarial training~\cite{Goodfellow2014Explaining, Rashid2022Adversarial, Xiong2023Aidtf, Gaber2025Robust}, out-of-distribution and anomaly detection~\cite{Chen2021Atom, Karunanayake2025Out}, input reconstruction~\cite{Gu2014Towards, Song2017Pixeldefend}, adversarial input detection~\cite{Lu2017Safetynet, Jiang2022Fgmd}, network  distillation~\cite{Papernot2016Distillation}, knowledge distillation~\cite{Levi2025Kdat}, and network verification~\cite{Katz2017Reluplex, Gopinath2017Deepsafe}, all aimed at enhancing model robustness. However, these defenses often incur trade-offs, including increased computational overhead, reduced accuracy on clean data, and limited adaptability to evolving threats~\cite{Pantelakis2023Adversarial, Han2021Evaluating}. Furthermore, their lack of interpretability and explainability hinders trustworthiness and accountability~\cite{Chander2025Toward}. To overcome these issues, we propose a novel, unsupervised adversarial detection model that enhances the robustness of IoT intrusion detection against adversarial attacks through SHapley Additive exPlanations (SHAP)~\cite{Lundberg2017Unified}-based attribution fingerprinting. The core intuition behind our approach is that adversarial attacks subtly distort a model’s internal reasoning, resulting in attribution patterns that deviate from those of clean inputs, despite appearing similar in the input space. These shifts are captured as SHAP fingerprints, which reflect anomalous feature attributions indicative of adversarial behavior. To detect such deviations, a deep learning-based autoencoder model is trained on SHAP vectors derived from clean data to learn the distribution of normal attribution patterns. During inference, inputs with high reconstruction error are flagged as adversarial network traffic, without the need for any labeled attack data. We summarize our {\bf key contributions} of this work as follows:
	\begin{itemize}
		\item Proposed a novel unsupervised SHAP attribution fingerprinting-based deep neural network model to enhance the robustness of IoT intrusion detection against adversarial attacks. The model learns distinct attribution patterns to detect adversarial inputs without needing labeled attack data. Our approach adds a defense layer to the IoT network by integrating with the IDS, improving its detection of adversarial attacks, and enhancing overall robustness.
        \item Developed an attribution fingerprinting generation pipeline that processes IoT traffic data, generates adversarial samples, and computes SHAP-based attribution fingerprints for both clean and adversarial inputs. These fingerprints serve as feature representations for training and evaluating the adversarial detection models, where training is performed on clean samples, and evaluation involves both clean and adversarial samples.
        \item Investigated the impact of different adversarial attacks on feature attributions using SHAP’s DeepExplainer method by identifying attack-specific feature's importance rank shifts. This rank shift captures variations between clean and adversarial samples, revealing distinct behavioral patterns induced by each attack. 
        \item Conducted extensive experiments for training and evaluation of the proposed model and its state-of-the-art baseline counterpart on a widely used IoT benchmark dataset. Evaluation results show that the proposed model significantly improves adversarial detection performance, robustness, and accuracy over the baseline.
        \item Enhanced the transparency, interpretability, and trustworthiness of the IoT intrusion detection model by leveraging explainable AI techniques for feature-level attribution analysis. This approach offers deeper insights into model decisions and strengthens user trust in the system.
	\end{itemize}
    
	The rest of this paper is organized as follows: Section \ref{sec:background_relatedwork} reviews the related work. Section \ref{sec:system-model} discusses the network and threat models considered in this work. Section \ref{sec:proposed-approach} introduces our proposed approach in terms of attribution fingerprint generation and adversarial intrusion detection model. Section \ref{sec:experiments} discusses the used dataset and experimental setup. Section~\ref{sec:adv-impact} discusses adversarial impact assessment with feature's rank using SHAP's attribution. Section~\ref{sec:evaluation}  presents model robustness and evaluation. Lastly, Section \ref{sec:conclusion} concludes this work and suggests future research directions.
	
	\section{Related Work} \label{sec:background_relatedwork}

    Explainable artificial intelligence (XAI) plays a crucial role in adversarial machine learning (AML) for enhancing the trustworthiness and robustness of AI and ML models~\cite{Chander2025Toward, Gafur2024Adversarial}. \citet{Noppel2024SoK} systematized the knowledge on how post-hoc explanation methods can serve as a foundation for developing more robust and explainable machine learning models. They further argued that if these methods are resilient to adversarial manipulation, they can be used as effective adversarial defense mechanisms. \citet{Siganos2023Explainable} proposed an ML and DL-based IDS for IoT networks that integrates explainability features. Their system leverages ML and DL models for the classification, while the SHAP~\cite{Lundberg2017Unified} is utilized to interpret and explain the model's decision-making process. Although the IDS model demonstrates promising performance in terms of classification accuracy, it does not consider adversarial attacks. \citet{Fidel2020When} proposed an adversarial detection method leveraging SHAP values from internal layers of a deep neural network to differentiate between normal and adversarial inputs. This approach demonstrates high detection accuracy, highlighting the effectiveness of SHAP-based explanations for identifying adversarial inputs. However, it relies entirely on image datasets, and its suitability is primarily limited to computer vision applications.

\citet{Wang2022Manda} proposed an adversarial example (AE) detection system that leverages both data manifold proximity and decision boundary analysis to identify adversarial traffic in ML-based IDS. Their experimental results confirmed that  AE attacks can effectively evade IDS with high success rates. To counter this, they identified common characteristics of successful AEs and designed a detector based on these insights. \citet{Gaber2025Robust} proposed a Generative Adversarial Networks (GAN)-based IDS for Internet of Flying Things (IoFT) that incorporates adversarial training to improve robustness against evolving threats. Similarly, \citet{Han2025DLR} proposed a detection and label recovery (DRL) framework that recovers ground-truth labels of detected adversarial examples using a separate detector and recovery components. The detector processes both legitimate and adversarial inputs, while the recovery module aims to infer the ground-truth label of detected adversarial examples. However, these approaches do not incorporate explainability features, limiting their transparency and interpretability.

\citet{Hewa2025Leveraging} proposed an explainable AI (XAI)-based framework for 6G network intrusion detection, incorporating real-time monitoring and concept drift detection against adaptive adversaries using the Wilcoxon signed-rank test~\cite{Rosner2006Wilcoxon}. This framework leverages XAI-derived feature importance to guide selective model retraining, thereby improving efficiency and robustness. In a related effort, \citet{Gummadi2024XAI-IoT} introduced an XAI-driven anomaly detection framework for IoT systems, utilizing both single and ensemble AI models, and applying multiple XAI methods to analyze feature importance across diverse models. While both frameworks enhance interpretability and detection accuracy, their evaluations were limited to conventional attacks without considering adversarial settings.

Despite recent advances in explainable AI (XAI) and adversarial machine learning, state-of-the-art deep learning-based network intrusion detection systems (NIDS) still exhibit several limitations. First, in terms of robustness, these systems often exhibit low accuracy and high false positive rates against adversarial attacks, a challenge that remains largely unaddressed in the context of IoT-based NIDS. Second, trustworthiness and transparency are undermined by the lack of explainability and interpretability. Third, there is a notable gap in enhancing adversarial robustness within IoT network environments, which present unique constraints such as limited computational resources, real-time communication demands, and heterogeneous device architectures and protocols. Additionally, IoT systems expose distinct attack surfaces, including wireless protocols (e.g., Zigbee, MQTT), physical sensor inputs, and edge-device firmware, all of which require tailored and context-aware defense strategies.

%Third, there is a notable gap in enhancing adversarial robustness specifically within IoT network environments, which pose unique constraints and attack surfaces that require tailored solutions.

In this paper, we propose a novel adversarial detection model that enhances the robustness of deep learning-based network intrusion detection systems (NIDS) in IoT networks through SHAP-based attribution fingerprinting. Our approach directly addresses the aforementioned limitations related to robustness, transparency, and IoT-specific applicability. To the best of our knowledge, this is the first systematic method that incorporates explainable AI into adversarial detection specifically tailored for IoT environments.

\section{System Model} \label{sec:system-model}
In this section, we describe the network model and threat model with the underlying assumptions made in this study. The network model defines the structure of the target IoT system and the deployed defense mechanisms. The threat model characterizes the adversarial capabilities, objectives, and types of attacks considered.

\subsection{Network Model}  \label{subsec:network-model}
We consider an Internet of Things (IoT) network comprising heterogeneous devices such as sensors, actuators, and smart appliances. The network configuration, including its underlying protocols, device types, and overall structure, is assumed to closely resemble the setup described in~\cite{EuclidesCICIoT2023}. Based on this setup, the network topology consists of 105 devices, and a total of 33 distinct attacks were executed. %These attacks are categorized into seven classes, including Distributed Denial of Service (DDoS), Denial of Service (DoS), Reconnaissance, Web-based, Brute-force, Spoofing, and Mirai. 

Each IoT device generates continuous network traffic, which is monitored by a centralized or edge-based NIDS. The traffic data is preprocessed into a structured feature set represented by input vectors \( x \in \mathbb{R}^d \), where \( d \) denotes the number of extracted features.The IDS is modeled as a binary classifier \( f_\theta: \mathbb{R}^d \rightarrow \{0, 1\} \), parameterized by \( \theta \), where the label \( y = 0 \) represents benign traffic and \( y = 1 \) represents malicious traffic. We consider this classifier implements a Feedforward Neural Network (FFNN)~\cite{Glorot2010Understanding, Goodfellow2016Deep}. 

%Suppose FFNN model consists of \( L \) layers. Each layer \( l \in \{1, 2, \dots, L\} \) is associated with a weight matrix \( \mathbf{W}^{(l)} \), a bias vector \( \mathbf{b}^{(l)} \), and an activation function \( \sigma^{(l)} \). The forward propagation through the network is recursively defined as:
%\begin{align}
%\mathbf{h}^{(0)} &= \mathbf{x} \quad \nonumber\\
%\mathbf{z}^{(l)} &= \mathbf{W}^{(l)} \mathbf{h}^{(l-1)} + \mathbf{b}^{(l)} \quad \nonumber \\
%\mathbf{h}^{(l)} &= \sigma^{(l)}\left(\mathbf{z}^{(l)}\right) \quad 
%\end{align}
%As this is a binary classification task, the output layer of the FFNN employs the sigmoid activation function to produce a probability score:
%\begin{align}
%\hat{y}  &= \frac{1}{1 + \exp\left(-\mathbf{z}^{(L)}\right)}
%\end{align}
%where, \( \hat{y} \in (0, 1) \) represents the predicted probability of the positive class.
%The model is trained by minimizing the binary cross-entropy loss function, which is defined as follows:  
%\begin{equation}
%\small
%\mathcal{L}(\theta) = -\frac{1}{N} \sum_{i=1}^{N} \left[ y_i \log(\hat{y}_i) + (1 - y_i) \log(1 - \hat{y}_i) \right]
%\end{equation}
%where, \( \hat{y}_i = f_\theta(\mathbf{x}_i) \in (0,1) \) denotes the predicted probability for input sample \( \mathbf{x}_i \), and \( y_i \in \{0,1\} \) is the corresponding ground-truth label, and \( \theta \) represents all learnable parameters of the network.

\subsection{Threat Model} \label{subsec:threat-model}
In this work, we consider a white-box adversarial attack model because SHAP relies on internal model information to compute precise feature attributions. In the white-box attack model, an adversarial has full access to the IoT intrusion detection model architecture and parameters \( \theta \)\cite{He2023Adversarial}. The adversary's goal is to craft adversarial examples \( x' = x + \delta \) such that the model output changes:
\vspace{-1mm}
    \[f_\theta(x') \ne f_\theta(x)\]

subject to a norm-bounded perturbation constraint:
\[
\|\delta\|_p \leq \epsilon
\]

where, \( \delta \in \mathbb{R}^d \) is a carefully crafted perturbation, \( \epsilon > 0 \) controls the attack strength, and \( \|\cdot\|_p \) denotes the \( \ell_p \)-norm. The adversary applies the following three well-established gradient-based attack methods to craft adversarial examples that aim to mislead or bypass the NIDS model during inference. These three attack methods include:
\begin{itemize}
    \item \textbf{Fast Gradient Sign Method (FGSM)} \cite{Goodfellow2014Explaining}: This attack perturbs the input in the direction of the gradient of the loss for an input, scaled by a small factor \( \epsilon \). The adversarial example is computed as:
   \begin{equation}
       x' = x + \epsilon \cdot \text{sign}(\nabla_x \mathcal{L}(f_\theta(x), y))
   \end{equation}
    
    where, \( \epsilon \) controls the perturbation magnitude and \( \text{sign}(\cdot) \) is the element-wise sign function.

    \item \textbf{Projected Gradient Descent (PGD)} \cite{Madry2017Towards}: This is an iterative version of FGSM that performs multiple small gradient ascent steps, projecting the perturbed input back onto the \( \ell_p \)-ball around the original input after each step. For each iteration \( t \), it can be defined as follows:
\begin{equation}
\resizebox{0.8\linewidth}{!}{$
     x^{(t+1)} = \Pi_{\mathcal{B}_\epsilon(x)} \left(x^{(t)} + \alpha \cdot \text{sign}(\nabla_x \mathcal{L}(f_\theta(x^{(t)}), y))\right)
     $}
\end{equation}
    where, \( \Pi_{\mathcal{B}_\epsilon(x)}(\cdot) \) denotes projection onto the \( \ell_p \)-ball of radius \( \epsilon \) centered at \( x \), and \( \alpha \) is a step size.

    \item \textbf{DeepFool} \cite{Moosavi2016Deepfool}: It is an iterative attack that aims to compute the minimal perturbation \( \delta \) required to change the classifier's decision. It assumes the classifier is approximately linear in a small neighborhood around the input \( x \). For a binary classifier \( f: \mathbb{R}^d \rightarrow \{-1, +1\} \), DeepFool finds the closest decision boundary by linearizing the classifier at point \( x \). The perturbation \( \delta \) is computed as:
\begin{equation}
    \delta = -\frac{f(x)}{\|\nabla f(x)\|_2^2} \nabla f(x)
\end{equation}

The adversarial example is then obtained by:
\[
x' = x + \delta
\]
The DeepFool attack iteratively updates the input with this minimal perturbation until the predicted label changes. This approach yields near-minimal adversarial perturbations that are often imperceptible but highly effective at misleading the classifier.
\end{itemize}

\section{Proposed Approach}\label{sec:proposed-approach}
In this section, we present our proposed approach that describes the attribution fingerprinting method and an adversarial intrusion detection model. The attribution fingerprinting method captures distinctive explanation patterns of input samples, while the detection model utilizes these patterns to effectively differentiate between clean and adversarial network traffic.

\subsection{Attribution Fingerprinting} \label{subsec:attribution-fingerprinting}
Attribution fingerprinting is a method that captures the distinctive patterns of feature attributions generated by explainable AI techniques such as SHAP~\cite{Lundberg2017Unified}, Local Interpretable Model-agnostic Explanations (LIME)~\cite{Ribeiro2016WhyShould} for a given input sample. In this work, we employ the SHAP method to generate attribution fingerprints that capture the contribution of each input feature to the model’s predictions. SHAP’s DeepExplainer is a model-specific method designed for deep neural networks that approximates Shapley values using the Deep Learning Important FeaTures (DeepLIFT)~\cite{Shrikumar2017Learning}.  

Let \( X = \{x_1, x_2, \dots, x_N\} \), \( x_i \in \mathbb{R}^M \), and \( Y = \{y_1, y_2, \dots, y_N\} \), \( y_i \in \{0, 1\} \), represents IoT traffic data, where each \( x_i \) denotes a network flow instance with \( M \) features, and corresponding label \( y_i \) indicates the class: \( y_i = 0 \) for benign and \( y_i = 1 \) for malicious sample, respectively. We compute the SHAP values using DeepExplainer for each input feature. Let $x_i$ be an $i^{th}$ input feature and its SHAP value is represented by $\phi(x_i)$ and computed as follows:
\begin{equation}
    \phi(x_i) = [\phi_1^{(i)}, \phi_2^{(i)}, \dots, \phi_M^{(i)}] \in \mathbb{R}^M
    \label{eq:shap}
\end{equation}
The model prediction for an input \( x_i \) can be approximated by decomposing it into the sum of a baseline value and feature attribution components, as shown below:
\begin{equation}
    f(x_i) \approx \phi_0 + \sum_{j=1}^M \phi_j^{(i)}
\end{equation}

where $\phi_0 = \mathbb{E}_{x'}[f(x')]$ is the expected output over a background distribution, and \(\phi_j^{(i)}\) denotes the attribution fingerprinting (i.e., SHAP value) of the \(j\)-th feature for input \(x_i\). We extract the fingerprinting for the malicious samples (i.e., class 1) and use them as the input vector for our proposed adversarial detection model. These vectors are computed as follows:
\begin{align}
    \label{eq:input-shap-values}
    Z &= \{ \phi(x_i) \mid y_i = 1 \} \nonumber \\
      &= \{z_1, z_2, \dots, z_K\}, \quad z_i \in \mathbb{R}^M
\end{align}
where each \( z_i \) represents the SHAP-based attribution vector corresponding to a malicious input sample \( x_i \).

\begin{figure*} 
\centering
\resizebox{0.85\textwidth}{!}{%
    {\includegraphics[width=.90\textwidth,keepaspectratio]{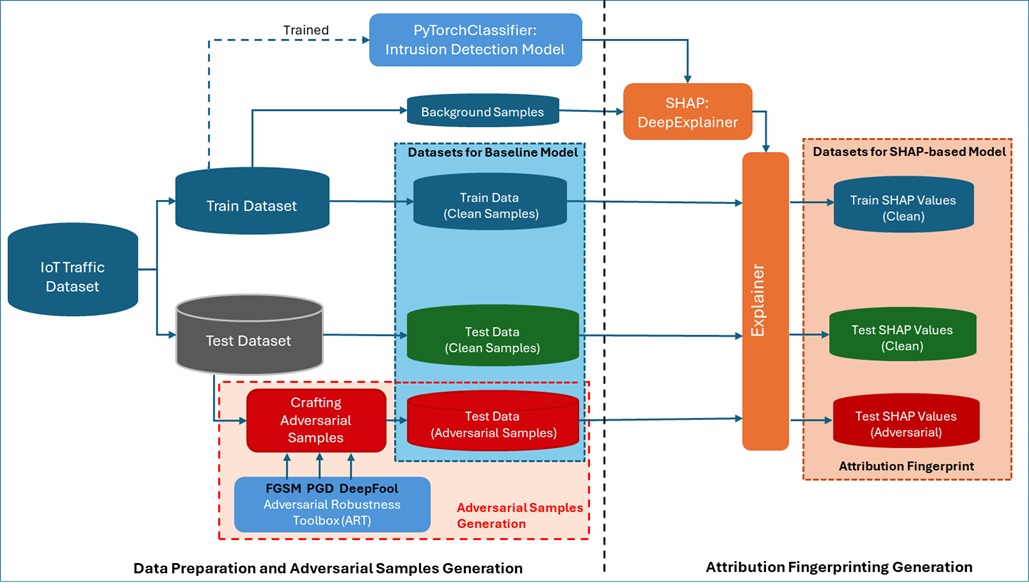}}
    }
\caption {\small Attribution Fingerprinting Generation.}
\label{fig:data-genearion-pipeline}
\end{figure*} 

\subsection{Adversarial Intrusion Detection Model} \label{Asussec:intrusion-detection-model}
Our proposed method leverages a deep autoencoder~\cite{Hinton2006Reducing} model as an unsupervised anomaly detector and trains it on SHAP attribution vectors of clean samples. The model learns the reconstruction patterns of clean data and flags adversarial inputs as anomalies when their reconstruction error deviates significantly from these learned patterns. 

Let \( A: \mathbb{R}^M \rightarrow \mathbb{R}^M \) denotes our proposed deep autoencoder model, which comprises an encoder \( E(\cdot) \), and a decoder \( D(\cdot) \) functions, each consisting of multiple hidden layers. These two components and their relationship with the hidden layers are defined as follows: 
\paragraph{Encoder.} The encoder maps the input \( \mathbf{z}_i \)  (SHAP attribution vector) to a lower-dimensional latent representation through a sequence of \( L \) nonlinear transformations. It is given as follows: 
\begin{align}
    \mathbf{h}_1 &= \sigma_1(W_1 \mathbf{z}_i + \mathbf{b}_1) \quad \nonumber \\
\mathbf{h}_2 &= \sigma_2(W_2 \mathbf{h}_1 + \mathbf{b}_2) \quad \nonumber \\  \quad  \quad \dots \nonumber \\
\mathbf{h}_L &= \sigma_L(W_L \mathbf{h}_{L-1} + \mathbf{b}_L) \nonumber
\end{align}
We denote this as:
\begin{equation}
\label{eq:encoder}
    \mathbf{h}_L = E(\mathbf{z}_i)
\end{equation}
where, \( \mathbf{h}_L \in \mathbb{R}^k \) is the latent representation of \( \mathbf{z}_i \), \( W_i \) and \( \mathbf{b}_i \) are learnable parameters (weight matrix and bias vector) of layer \( i \), and \( \sigma_i(\cdot) \) is a nonlinear activation function.
\paragraph{Decoder.}
The decoder reconstructs the input from the latent representation of $\mathbf{h}_L$. Decoder produces output of the model, which is denoted by  $\hat{\mathbf{z}}_i$ and  obtained as follows:
\begin{equation}
\label{eq:decoder}
    \hat{\mathbf{z}}_i = D(\mathbf{h}_L)
\end{equation}

The model learns the underlying distribution of clean SHAP attribution vectors by minimizing the reconstruction loss during training. Specifically, the objective is to minimize the mean squared reconstruction error over a dataset of \( K \) clean samples. The loss function is defined as:
\begin{equation}
    \mathcal{L} = \frac{1}{K} \sum_{i=1}^{K} \left\| \mathbf{z}_i - \hat{\mathbf{z}}_i \right\|_2^2
\end{equation}

At inference, when a new input sample is received, the reconstruction error of its corresponding SHAP attribution vector is computed using the trained autoencoder and compared this error against a predefined threshold ($\tau$), and an anomaly decision is made accordingly. For a given new input sample $x^*$, the adversarial (i.e., anomaly) detection involves the following computation steps: 
\begin{enumerate}
    \item SHAP vector: $z^* = \phi(x^*)$
    \item Reconstruction: $\hat{z}^* = A(z^*)$
    \item Reconstruction error: $s(x^*) = \| z^* - \hat{z}^* \|_2^2$
    \item Decision: \[
\hat{y}^* =
\begin{cases}
\text{Clean}, & \text{if } s(x^*) \leq \tau  \\
\text{Adversarial}, & \text{if } s(x^*) > \tau  
\end{cases}
\]
\end{enumerate}
where, \( \hat{y}^* \) is the predicted label for the input \( x^* \). The threshold \( \tau \) can be determined from clean samples of validation dataset using standard statistical techniques such as $95^{th}$ or $99^{th}$ percentile of reconstruction errors, or by applying the empirical \( 2\sigma \) or \( 3\sigma \) rule, assuming that the reconstruction errors follow a Gaussian distribution.

\section{Experimental Setup} \label{sec:experiments}
In this section, we describe the datasets and experimental setup used in this study. It includes a dataset description, data reprocessing, adversarial samples generation, and attribution fingerprinting generation.

\subsection{Dataset} \label{subsec:dataset}
In this work, we use the CIC-IoT2023~\cite{EuclidesCICIoT2023} dataset to train and evaluate our proposed model. This large-scale, real-world benchmark is widely adopted in the IoT security research community. It contains network traffic captured from 105 actual IoT devices within a controlled testbed environment and includes both benign and attacks such as DoS, DDoS, reconnaissance, web-based exploits, brute force, spoofing, and Mirai botnet attacks.

We utilized a subset of the CIC-IoT2023 dataset by selecting and merging the first 15 CSV files to create a binary classification dataset, grouping all attack types under the malicious class. From this dataset, we derived inputs for adversarial sample generation and SHAP-based attribution fingerprinting. Fig.~\ref{fig:data-genearion-pipeline} illustrates the overall pipeline for generating adversarial samples and SHAP-based attribution fingerprints. The process begins by splitting the dataset into training and testing subsets. The training set is exclusively used to train all models, including the reference NIDS classifier and the proposed SHAP-based adversarial detection model. The test set is reserved solely for evaluation purposes.

The data generation pipeline consists of two main steps: adversarial sample generation and SHAP-based attribution fingerprinting. First, adversarial samples are generated from the test set using the Adversarial Robustness Toolbox (ART)~\cite{Nicolae2018Adversarial, ART2025}, stored them, and then passed through a SHAP explainer to compute their attribution fingerprints. The second step involves the generation of SHAP-based fingerprints using the SHAP framework\cite{Lundberg2017Unified}. To generate attribution fingerprints, we first created an explainer using SHAP’s DeepExplainer~\cite{Lundberg2017Unified}, with a background dataset sampled from the clean training set and the trained NIDS model. This explainer was then applied to both clean and adversarial samples to produce their attribution fingerprints. These fingerprints are stored and used for feature importance analysis, assessing adversarial impact, and training and evaluating our proposed detection model.
\vspace{-1.5mm}
\begin{table*}
\centering
\caption{SHAP Value, Feature's Rank, and Rank Shift under Adversarial Attacks}
\resizebox{0.85\textwidth}{!}{%
\begin{tabular}{l|c|cccc|cccc|ccc}
\toprule
\multirow{2}{*}{\textbf{Feature Name}} & \multirow{2}{*}{\textbf{Index}} & \multicolumn{4}{c|}{\textbf{SHAP Values}} & \multicolumn{4}{c|}{\textbf{Feature's Rank}} & \multicolumn{3}{c}{\textbf{Feature's Rank Shift}} \\
\cmidrule(lr){3-6} \cmidrule(lr){7-10} \cmidrule(lr){11-13}
 & & Clean & FGSM & PGD & DeepFool & Clean & FGSM & PGD & DeepFool & FGSM & PGD & DeepFool \\
\midrule
Number & 37 & 1 & 1 & 1 & 1 & 1 & 1 & 1 & 1 & 0 & 0 & 0 \\
ack\_flag\_number & 8 & 0.2782 & 0.1667 & 0.2354 & 0.1667 & 2 & 11 & 6 & 11 & 9 & 4 & 9 \\
Header\_Length & 0 & 0.1770 & 0.3948 & 0.3297 & 0.3948 & 3 & 4 & 3 & 4 & 1 & 0 & 1 \\
fin\_flag\_number & 4 & 0.1722 & 0.1795 & 0.1473 & 0.1795 & 4 & 8 & 13 & 8 & 4 & 9 & 4 \\
ack\_count & 11 & 0.1274 & 0.1411 & 0.2530 & 0.1411 & 5 & 13 & 5 & 13 & 8 & 0 & 8 \\
syn\_flag\_number & 5 & 0.1235 & 0.1644 & 0.09194 & 0.1644 & 6 & 12 & 17 & 12 & 6 & 11 & 6 \\
TCP & 22 & 0.1125 & 0.3166 & 0.3302 & 0.3166 & 7 & 5 & 2 & 5 & 2 & 5 & 2 \\
Max & 32 & 0.1050 & 0.1019 & 0.2164 & 0.1019 & 8 & 16 & 7 & 16 & 8 & 1 & 8 \\
rst\_flag\_number & 6 & 0.0904 & 0.0803 & 0.0695 & 0.0803 & 9 & 17 & 21 & 17 & 8 & 12 & 8 \\
UDP & 23 & 0.0726 & 0.1802 & 0.1905 & 0.1802 & 10 & 7 & 8 & 7 & 3 & 2 & 3 \\
HTTPS & 16 & 0.0689 & 0.1149 & 0.1141 & 0.1149 & 11 & 15 & 15 & 15 & 4 & 4 & 4 \\
Time\_To\_Live & 2 & 0.0640 & 0.1697 & 0.1260 & 0.1697 & 12 & 10 & 14 & 10 & 2 & 2 & 2 \\
syn\_count & 12 & 0.0607 & 0.1199 & 0.0770 & 0.1199 & 13 & 14 & 18 & 14 & 1 & 5 & 1 \\
Rate & 3 & 0.0448 & 0.0387 & 0.0564 & 0.0387 & 14 & 26 & 24 & 26 & 12 & 10 & 12 \\
Tot sum & 30 & 0.0395 & 0.0755 & 0.1606 & 0.0755 & 15 & 18 & 12 & 18 & 3 & 3 & 3 \\
fin\_count & 13 & 0.0362 & 0.0668 & 0.0554 & 0.0668 & 16 & 19 & 25 & 19 & 3 & 9 & 3 \\
ICMP & 26 & 0.0319 & 0.0504 & 0.0585 & 0.0504 & 17 & 22 & 23 & 22 & 5 & 6 & 5 \\
psh\_flag\_number & 7 & 0.0303 & 0.1755 & 0.1014 & 0.1755 & 18 & 9 & 16 & 9 & 9 & 2 & 9 \\
Std & 34 & 0.0301 & 0.0282 & 0.0347 & 0.0282 & 19 & 29 & 30 & 29 & 10 & 11 & 10 \\
Tot size & 35 & 0.0266 & 0.3996 & 0.1896 & 0.3996 & 20 & 3 & 9 & 3 & 17 & 11 & 17 \\
rst\_count & 14 & 0.0251 & 0.0450 & 0.0349 & 0.0450 & 21 & 25 & 29 & 25 & 4 & 8 & 4 \\
AVG & 33 & 0.0250 & 0.2916 & 0.1644 & 0.2916 & 22 & 6 & 11 & 6 & 16 & 11 & 16 \\
HTTP & 15 & 0.0143 & 0.0602 & 0.0494 & 0.0602 & 23 & 21 & 26 & 21 & 2 & 3 & 2 \\
Protocol Type & 1 & 0.0139 & 0.0155 & 0.0129 & 0.0155 & 24 & 34 & 35 & 34 & 10 & 11 & 10 \\
DNS & 17 & 0.0136 & 0.0201 & 0.0374 & 0.0201 & 25 & 32 & 28 & 32 & 7 & 3 & 7 \\
LLC & 29 & 0.0054 & 0.0458 & 0.0451 & 0.0458 & 26 & 24 & 27 & 24 & 2 & 1 & 2 \\
ARP & 25 & 0.0032 & 0.0066 & 0.0032 & 0.0066 & 27 & 36 & 38 & 36 & 9 & 11 & 9 \\
IPv & 28 & 0.0029 & 0.0274 & 0.0299 & 0.0274 & 28 & 30 & 31 & 30 & 2 & 3 & 2 \\
Min & 31 & 0.0026 & 0.0083 & 0.0265 & 0.0083 & 29 & 35 & 32 & 35 & 6 & 3 & 6 \\
Variance & 38 & 0.0018 & 0.0335 & 0.0 & 0.0335 & 30 & 28 & 39 & 28 & 2 & 9 & 2 \\
cwr\_flag\_number & 10 & 0.0015 & 0.0234 & 0.0188 & 0.0234 & 31 & 31 & 33 & 31 & 0 & 2 & 0 \\
SSH & 20 & 0.0014 & 0.0040 & 0.0668 & 0.0040 & 32 & 37 & 22 & 37 & 5 & 10 & 5 \\
IAT & 36 & 0.0011 & 0.4972 & 0.2682 & 0.4972 & 33 & 2 & 4 & 2 & 31 & 29 & 31 \\
ece\_flag\_number & 9 & 0.0009 & 0.0343 & 0.0169 & 0.0343 & 34 & 27 & 34 & 27 & 7 & 0 & 7 \\
DHCP & 24 & 0.0007 & 0.0039 & 0.0074 & 0.0039 & 35 & 38 & 36 & 38 & 3 & 1 & 3 \\
IGMP & 27 & 0.0002 & 0.0500 & 0.0748 & 0.0500 & 36 & 23 & 19 & 23 & 13 & 17 & 13 \\
Telnet & 18 & 0.0002 & 0.0188 & 0.1774 & 0.0188 & 37 & 33 & 10 & 33 & 4 & 27 & 4 \\
IRC & 21 & 9.97E-05 & 0.0625 & 0.0719 & 0.0625 & 38 & 20 & 20 & 20 & 18 & 18 & 18 \\
SMTP & 19 & 0.0 & 0.0 & 0.0068 & 0.0& 39 & 39 & 37 & 39 & 0 & 2 & 0 \\
\bottomrule
\end{tabular}%
}
\label{tab:feature-ranking}
\end{table*}
\begin{figure}
 {\includegraphics[width=.5\textwidth,keepaspectratio]{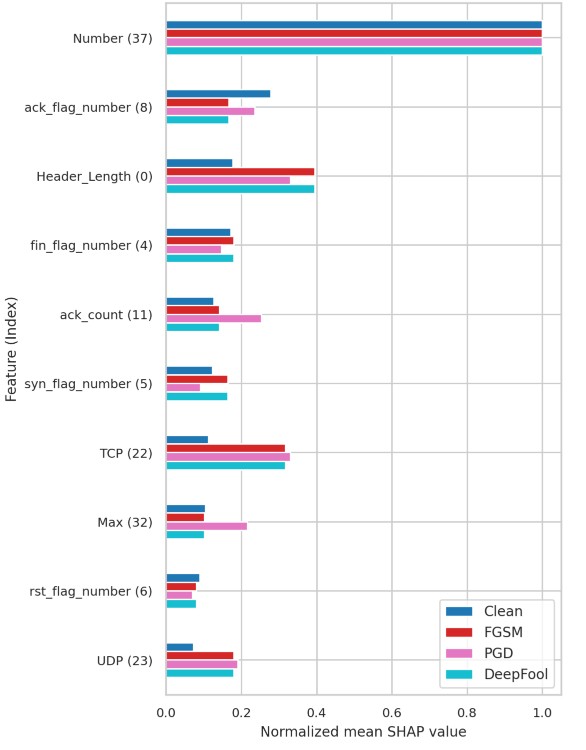}}
\caption {\small Comparing Top-10 Features of Clean and Attack Samples Based on SHAP-Attribution.}
\label{fig:top-10-features}
\end{figure}    
\subsection{Experimental Setup} \label{subsec:experimental-setup}

We set up the experimental environment based on the network and threat model discussed in Section~\ref{sec:system-model} and conducted several experiments to develop and evaluate our SHAP-based adversarial intrusion detection model. First, we developed and trained a reference IoT NIDS model to support SHAP attribution generation. Second, we crafted adversarial samples using ART~\cite{Nicolae2018Adversarial} for three representative attacks such as FGSM~\cite{Goodfellow2014Explaining}, PGD~\cite{Madry2017Towards}, and DeepFool~\cite{Moosavi2016Deepfool}. These adversarial samples are used to assess how well the model can detect subtle, malicious perturbations designed to bypass our detector. Third, we computed attribution fingerprints for both clean and adversarial data using SHAP framework~\cite{Lundberg2017Unified}. These attribution fingerprint data are primarily used in our model for both training and evaluation.  Fourth, we performed feature importance analysis and attribution rank shift comparisons to examine how attacks influence model explanations. All experiments were performed on a Tesla V100 GPU system using a PyTorch-based deep learning environment. For the comparative performance analysis, we trained the following two detector models: 
   \begin{itemize}
   \item {\bf SHAP-based Model}: This is an implementation of the unsupervised deep autoencoder model described in Section~\ref{Asussec:intrusion-detection-model}. It is trained using our proposed approach with SHAP attribution fingerprints. Its robustness is evaluated against adversarial attacks and compared with a state-of-the-art adversarial defense method~\cite{He2023Adversarial}. 
   
   \item {\bf Adversarially Trained Model}: This model incorporates an adversarial defense strategy based on adversarial training~\cite{Madry2017Towards}. It is trained using adversarial examples generated with the FGSM attack~\cite{Goodfellow2014Explaining,Nicolae2018Adversarial}. This serves as a benchmark defense method, enabling a comparative evaluation with our proposed approach.
\end{itemize}

   As both models are unsupervised deep autoencoders, they are trained exclusively on clean data and their corresponding SHAP-based attributions. We evaluated their effectiveness with standard classification and adversarial robustness metrics. This dual-model setup enables a systematic evaluation of the effectiveness of our attribution-guided learning in detecting adversarial inputs.

\section{Adversarial Impact Assessment with Feature's Rank} \label{sec:adv-impact}

In this section, we present an adversarial impact analysis using SHAP-based feature rankings. We examine how adversarial perturbations alter feature importance and compute rank shifts to identify features that distinctly separate clean and adversarial inputs. This analysis also highlights which features are more vulnerable or consistently robust under attack.

\subsection{Feature Importance Analysis} \label{subsec:feature-imp}
To assess the impact of adversarial perturbations, we first compare the SHAP-based feature importance for clean and adversarial samples. We comparatively analyzed the feature importance of both clean and each attack sample. Fig.~\ref{fig:top-10-features} shows the top 10 most influential features plotting SHAP values for clean data and adversarial samples generated using FGSM~\cite{Goodfellow2014Explaining}, PGD~\cite{Madry2017Towards}, and DeepFool~\cite{Moosavi2016Deepfool} attacks. This comparative analysis provides critical insights into how adversarial perturbations affect feature attribution. Notably, the Number feature (index 37) remains stable across all sample types, suggesting it is robust to adversarial manipulation. In contrast, features such as Header\_Length, TCP, and UDP demonstrate marked shifts in importance between clean and adversarial inputs, highlighting their susceptibility to attack-induced distortions. Additionally, the Ack\_Feature shows a relatively diminished importance in the presence of adversarial samples. The SHAP plots also identify attack-specific feature behavio. For example, Ack\_Count and Max features exhibit notably higher importance in PGD samples, while their influence is reduced under FGSM, DeepFool, and clean conditions.
\vspace{-1mm}
\begin{figure*} 
    \centering
    \scalebox{0.90}{ % 
    \resizebox{0.65\textwidth}{!}{%
    \begin{tabular}{cc}
        \subfloat[SHAP-based: Clean vs. FGSM]{%
            \includegraphics[ width=0.45\textwidth,keepaspectratio]{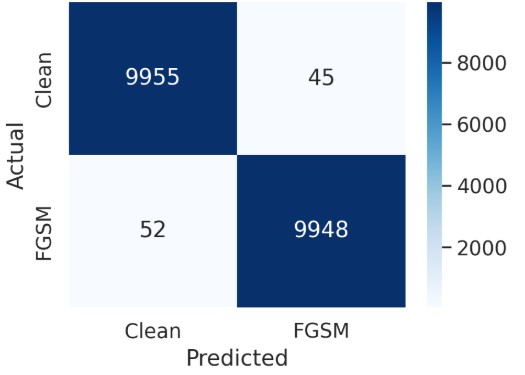}} &
        \subfloat[Adversarially Trained: Clean vs. FGSM]{%
            \includegraphics[ width=0.45\textwidth,keepaspectratio]{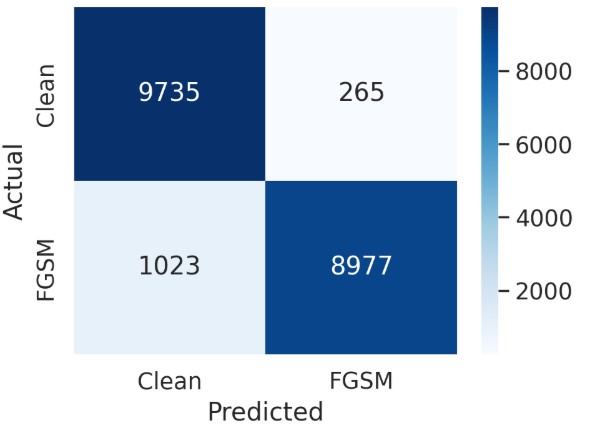}} \\
        
        \subfloat[SHAP-based: Clean vs. PGD]{%
            \includegraphics[width=0.45\textwidth]{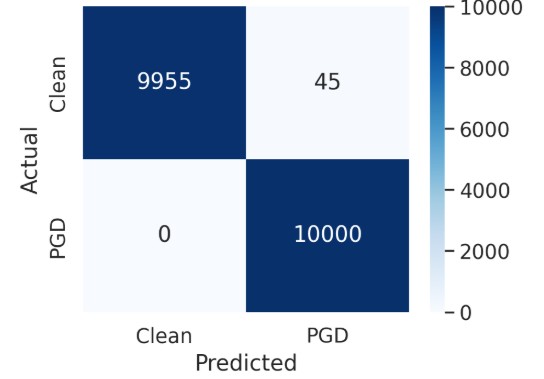}} &
        \subfloat[Adversarially Trained: Clean vs. PGD]{%
            \includegraphics[ width=0.45\textwidth,keepaspectratio]{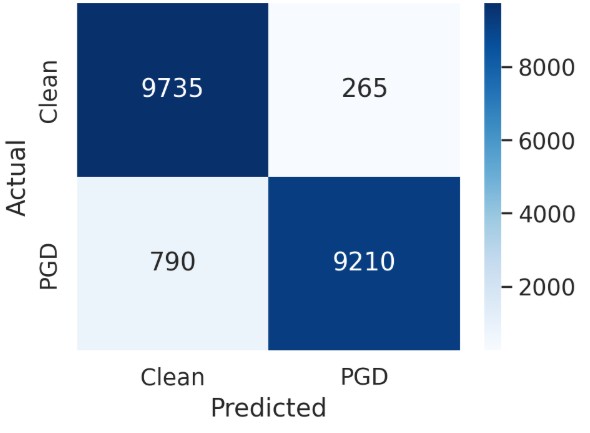}} \\
       
        \subfloat[SHAP-based: Clean vs. DeepFool]{%
            \includegraphics[ width=0.45\textwidth,keepaspectratio]{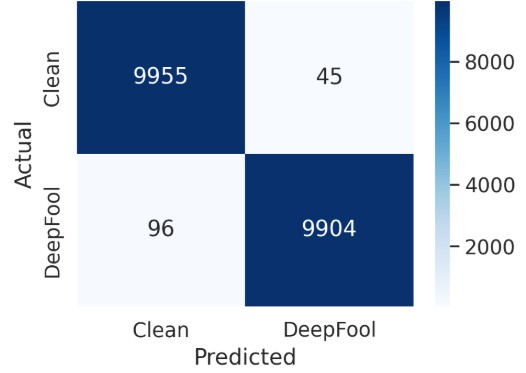}} &
        \subfloat[Adversarially Trained: Clean vs. DeepFool]{%
            \includegraphics[width=0.45\textwidth]{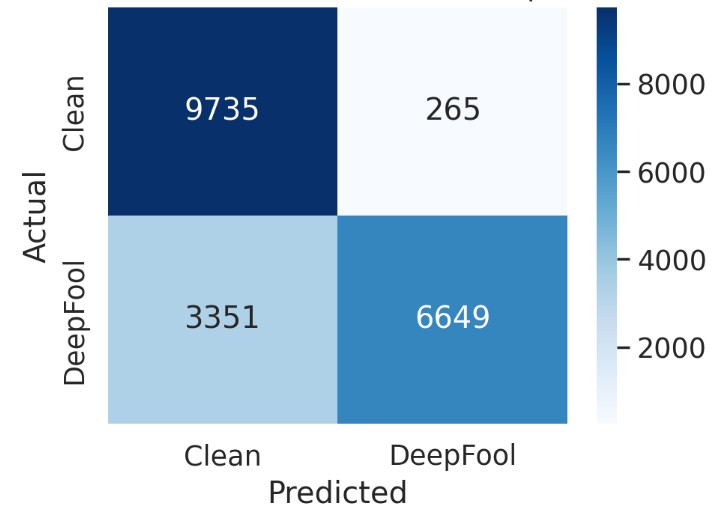}} 
    \end{tabular}%
    }
    }
    \caption{\small Comparison of adversarial detection performance of our proposed SHAP-based model against the adversarially trained model with different attack scenarios: (a–b) FGSM, (c–d) PGD, and (e–f) DeepFool.}
    \label{fig:confusion-mat-shap-model-vs-baseline}
\end{figure*}

\subsection{Feature's Rank Shift Analysis} \label{subsec:shap-values analysis}
We further investigated the impact of attacks on each feature's importance with SHAP-attribution by computing their rank shifts relative to the clean input samples.  It provides key insight into how feature attributions change under adversarial perturbations, revealing which features become more or less influential and indicating the model’s vulnerability or resilience to such attacks.

Table~\ref{tab:feature-ranking} presents the computed SHAP values for each feature, along with their importance ranks and the corresponding rank shifts under adversarial attacks relative to clean samples. Analysis of SHAP  values quantifies the contribution of each feature to the model’s prediction. Higher SHAP values indicate more influence on the model's output. A feature whose SHAP value significantly changes under attack may indicate the model's vulnerability or instability in that dimension. Feature ranking allows us to assess how the relative importance of features shifts under adversarial attacks. Each feature's rank shift shows how much of its importance has changed relative to the clean. The feature 'Number' holds a consistent rank of 1 across all attacks and clean, with a rank shift of 0, indicating strong robustness and stability under adversarial perturbations. In contrast, features such as 'IAT', 'IRC', and 'Total size' exhibit large rank shifts, showing higher sensitivity to adversarial manipulation and reduced reliability in adversarial settings. A large rank shift shows that adversarial inputs significantly alter the model’s attribution of feature importance and potentially mislead the model to rely on less relevant features in its decision-making process. This can be indicative of successful adversarial manipulation, potentially compromising the model's interpretability and robustness.
%Table~\ref{tab:feature-ranking} presents results of a comprehensive analysis of how SHAP values, feature importance rankings, and their rank shifts behave under different adversarial attacks compared to clean input data. 

\section{Model Robustness \& Evaluation} \label{sec:evaluation}
In this section, we present a comparative and quantitative evaluation of the proposed SHAP-based detection model against a state-of-the-art adversarial defense method (i.e., adversarial training) using standard benchmark classification, discrimination, and robustness metrics. We first describe the evaluation metrics, followed by results and analysis, including confusion matrices, metric-based performance comparisons, and reconstruction error distribution analysis. This analysis demonstrates the effectiveness of our proposed model and resilience under adversarial settings.

%Performance comparision table
\begin{table*} [!h]
\centering
\caption{\small Comparing classification performance of our proposed SHAP-based model against state-of-the-art adversarially trained benchmark model on three adversarial attacks via evaluation metrics.}
\resizebox{0.98\textwidth}{!}{%
\begin{tabular}{llcccccccccccccc}
\toprule
\textbf{Models} & \textbf{Attacks} & \textbf{Accuracy} & \textbf{Precision} & \textbf{Recall} & \textbf{F1} & \textbf{AUC} & \textbf{AP} & \textbf{TNR} & \textbf{NPV} & \textbf{FPR} & \textbf{FNR} & \textbf{TP} & \textbf{TN} & \textbf{FP} & \textbf{FN} \\
%\midrule
%\multirow{3}{*}{Baseline Model}
 %& FGSM & 0.9616 & 0.9886 & 0.9340 & 0.9605 & 0.9970 & 0.9954 & 0.9892 & 0.9375 & 0.0108 & 0.0660 & 9340 & 9892 & 108 & 660 \\
 %& PGD  & 0.9899 & 0.9892 & 0.9906 & 0.9899 & 0.9986 & 0.9975 & 0.9892 & 0.9906 & 0.0108 & 0.0094 & 9906 & 9892 & 108 & 94  \\
 %& DeepFool & 0.8300 & 0.9842 & 0.6707 & 0.7977 & 0.9876 & 0.9853 & 0.9892 & 0.7502 & 0.0108 & 0.3293 & 6707 & 9892 & 108 & 3293 \\
\midrule
\multirow{3}{*}{\parbox{2cm}{Our SHAP-based \\ Model}}
 & FGSM & 0.9952 & 0.9955 & 0.9948 & 0.9951 & 0.9987 & 0.9972 & 0.9955 & 0.9948 & 0.0045 & 0.0052 & 9948 & 9955 & 45 & 52 \\
 & PGD  & 0.9978 & 0.9955 & 1.0000 & 0.9978 & 0.9993 & 0.9981 & 0.9955 & 1.0000 & 0.0045 & 0.0000 & 10000 & 9955 & 45 & 0 \\
 & DeepFool & 0.9930 & 0.9955 & 0.9904 & 0.9929 & 0.9986 & 0.9972 & 0.9955 & 0.9904 & 0.0045 & 0.0096 & 9904 & 9955 & 45 & 96 \\
 \midrule
\multirow{3}{*}{\parbox{2cm}{Adversarially Trained Model}}
 & FGSM & 0.9356 &  0.9713 &        0.8977 &  0.9330 &  0.9898 &  0.9856 &             0.9735 &  0.9049 &  0.0265 &  0.1023 &  8977.0 & 9735.0 &  265.0 & 1023.0 \\
 & PGD  & 0.9472   & 0.9720        & 0.9210  &0.9458  &0.9947  & 0.9921            & 0.9735  & 0.9249 & 0.0265 & 0.0790  & 9210.0  & 9735.0 & 265.0  & 790.0             \\
 & DeepFool & 0.81920  & 0.9616        & 0.6649  & 0.7862  & 0.9691 & 0.9626             & 0.9735  & 0.7439  & 0.0265  & 0.3351  & 6649.0  & 9735.0  & 265.0  & 3351.0              \\
\bottomrule
\end{tabular}%
\label{tab:performance-comparision}
}
\end{table*}
%& 0.9735 & 0.8977 &0.1023
%& 0.9735   & 0.9210 &0.0790
%& 0.9735  & 0.6649 &0.3351 
\normalsize
\vspace{-1mm}
\subsection{Evaluation Metrics} \label{subsec:metrics}

In adversarial detection, evaluation metrics quantify a model’s ability to distinguish clean from adversarial inputs. We compared the performance of our SHAP-based model against an adversarially trained neural network model with an identical architecture trained on adversarial examples. The evaluation employed a comprehensive set of classification and robustness metrics. The classification metrics include accuracy,  precision, recall (true positive rate), F1-score, ROC AUC, average precision (AP), specificity (true negative rate), negative predictive value (NPV), false positive rate (FPR), false negative rate (FNR), true positive (TP), true negative (TN), false positive (FP), and False negative (FN). These metrics are defined by the following equations:

\footnotesize
\vspace{-1.5mm}
\begin{equation}
\text{Accuracy} = \frac{TP + TN}{TP + TN + FP + FN}
\label{eq:accuracy}
\end{equation}
\vspace{-2mm}
\begin{equation}
\text{Precision} = \frac{TP}{TP + FP}
\label{eq:precision}
\end{equation}
\vspace{-3mm}
\begin{equation}
\text{Recall} = \frac{TP}{TP + FN}
\label{eq:recall}
\end{equation}
\vspace{-3mm}
\begin{equation}
\text{F1-Score} = 2 \cdot \frac{\text{Precision} \cdot \text{Recall}}{\text{Precision} + \text{Recall}}
\label{eq:f1}
\end{equation}
\vspace{-3mm}
\begin{equation}
\text{ROC AUC} = \int_0^1 \text{TPR}(FPR) \, d(\text{FPR})
\label{eq:roc_auc}
\end{equation}
\vspace{-2mm}
\begin{equation}
\text{AP} = \sum_n (R_n - R_{n-1}) \cdot P_n
\label{eq:ap}
\end{equation}
\vspace{-2mm}
\begin{equation}
\text{Specificity} = \frac{TN}{TN + FP}
\label{eq:specificity}
\end{equation}
\vspace{-2mm}
\begin{equation}
\text{NPV} = \frac{TN}{TN + FN}
\label{eq:npv}
\end{equation}

\begin{equation}
\text{FPR} = \frac{FP}{FP + TN}
\label{eq:fpr}
\end{equation}

\begin{equation}
\text{FNR} = \frac{FN}{FN + TP}
\label{eq:fnr}
\end{equation}

%The accuracy metric quantifies the overall rate of correct classifications of the model. Precision measures the proportion of predicted adversarial samples that are truly adversarial, reflecting false alarm rates. Recall (TPR) captures the model’s ability to detect actual adversarial instances. The F1-score balances precision and recall via their harmonic mean. ROC AUC evaluates the model’s discrimination ability across all thresholds, while AP summarizes precision-recall trade-offs over varying confidence levels. Specificity (TNR) assesses the correct identification of clean inputs, and NPV measures the reliability of predicted clean labels. FPR and FNR capture the rates of misclassified clean and adversarial samples, respectively. TP, TN, FP, and FN collectively characterize the model’s detection performance.
\normalsize
We further evaluate the robustness of our proposed approach using Clean Accuracy (CA), Adversarial Accuracy (AA), and Attack Success Rate (ASR) metrics. 
CA measures the proportion of clean inputs classified correctly, while AA measures the proportion of adversarial inputs classified correctly.  ASR metric measures the effectiveness of adversarial attacks as the proportion of adversarial inputs that cause misclassification, providing a direct measure of attack success. These metrics are computed as follows:
\vspace{-1mm}
\small
\begin{align}
\centering
\mathrm{CA} &= \frac{\#\{\text{correctly classified clean samples}\}}{\#\{\text{total clean samples}\}} \\
\mathrm{AA} &= \frac{\#\{\text{correctly classified adversarial samples}\}}{\#\{\text{total adversarial samples}\}} \\
\mathrm{ASR} &= \frac{\#\{\text{misclassified adversarial samples}\}}{\#\{\text{total adversarial samples}\}} 
\end{align}
\normalsize

%Results and analysis section
\subsection{Results \& Analysis}
\label{subsec:result-analysis}

The evaluation results are analyzed using confusion matrices, metric-based classification and robustness comparisons, and reconstruction error distribution to assess the model’s performance and discriminative capability.

%Confusion matrix
\subsubsection{Confusion Matrix}
Fig.~\ref{fig:confusion-mat-shap-model-vs-baseline} shows the confusion matrices comparing the prediction results of our SHAP-based model against an adversarially trained baseline model and evaluated on clean and adversarial inputs. Each confusion matrix illustrates the model’s ability to distinguish clean inputs from adversarial samples. The confusion matrix analysis highlights the effectiveness of our model in accurately distinguishing between clean and adversarial samples. It achieves a low false positive rate, misclassifying only 45 out of 10,000 clean samples as adversarial, compared to 265 by the adversarially trained baseline model. This indicates improved precision and minimal disruption to clean inputs. In terms of false negatives, our model consistently outperforms the baseline across all attack types. Under FGSM, it misclassifies only 52 adversarial samples, whereas the baseline fails on 1,023. For the more challenging PGD attack, our model achieves zero false negatives, compared to 790 by the baseline. Similarly, under DeepFool, our model misclassifies 96 samples, which is significantly fewer than the 3,351 misclassified by the baseline. These results demonstrate that our model offers superior robustness and sensitivity across a range of adversarial threats while maintaining high accuracy on clean data.

%Matric-based comparision
\subsubsection{Metric-based Comparative Performance Analysis} \label{subsec:robustness-metrics-results}
We comparatively assess the classification and robustness, and discrimination capability of our model against a baseline counterpart for each attack type using the state-of-the-art evaluation metrics. 

\paragraph{Classification Metrics} The classification metrics reflect the model’s predictive performance across clean and adversarial conditions. Table~\ref{tab:performance-comparision} presents the overall performance comparison of the models using the classification metrics. Results show that our SHAP-based model performs very well against all attacks. For example, under FGSM, our SHAP-based model achieves accuracy = 0.9952, precision = 0.9955, recall = 0.9948, and F1-score = 0.9951, all significantly higher than the adversarially trained baseline model (accuracy = 0.9356, F1 = 0.9330). A low false positive rate (FPR) indicates the model rarely flags clean data as adversarial, preserving usability, while a low false negative rate (FNR) shows strong attack detection. Under DeepFool attack, our SHAP-based model reduces FNR dramatically to 0.0096 (96 false negatives), compared to the baseline’s much higher 0.3351 (3,351 false negatives). Similarly, the SHAP-based model maintains an FPR of 0.0045, less than half the benchmark baseline’s 0.0265, indicating better protection against unnecessary alerts on clean data. These values indicate our model has superior detection accuracy with minimal trade-off between precision and recall. Similar improvements are observed under PGD and DeepFool, where the SHAP-based model maintains F1-scores above 0.99, confirming its consistent and balanced adversarial detection capability.

\paragraph{Discrimination metrics} Discrimination metrics measure how effectively the model separates adversarial samples from clean samples with given thresholds $\tau$, reflecting the confidence and reliability of detection. A higher AUC and Average Precision (AP) indicate better overall separability regardless of threshold choice. For PGD attacks, our SHAP-based model attains an AUC of 0.9993 and AP of 0.9981, slightly surpassing the baseline’s AUC of 0.9898 and AP of 0.9856. These values demonstrate that our approach distinguishes attacks more effectively, reinforcing its robustness and reliability in adversarial scenarios.

\paragraph{Robustness Metrics} Robustness metrics evaluate a model’s ability to accurately classify clean samples while effectively detecting adversarial inputs, highlighting resilience to misclassification. Table~\ref{tab:performance-comparision-robustness} shows a comparative analysis of our SHAP-based approach against a state-of-the-art adversarial defense method (i.e., adversarial training) using standard robustness metrics. Our SHAP-based model maintains consistently high clean accuracy (CA $\approx$ 0.9955) and minimal degradation in adversarial accuracy (AA $\geq$ 0.9904) across all attack types, resulting in very low attack success rates (ASR $\leq$ 0.0096). In contrast, the adversarially trained model exhibits lower CA ($\approx$ 0.9735) and experiences substantial drops in AA, particularly under DeepFool attacks (AA = 0.6649), leading to significantly higher ASR (ASR = 0.3351). These findings demonstrate that our SHAP-based defense provides superior robustness against diverse adversarial perturbations while preserving high fidelity on clean inputs.
\vspace{-1mm}
\begin{table} 
\centering
\caption{\small Comparison of our SHAP-based approach and state-of-the-art defense via robustness evaluation metrics.}
\resizebox{0.48\textwidth}{!}{%
\begin{tabular}{llcccccccccccccc}
\toprule
\textbf{Models} & \textbf{Attacks} & \textbf{CA} & \textbf{AA} & \textbf{ASR} \\
\midrule
\multirow{3}{*}{\parbox{2cm}{Our SHAP-based \\ Model}}
 & FGSM      & 0.9955 & 0.9948 & 0.0052 \\
& PGD       & 0.9955 & 1.0000 & 0.0000 \\
& DeepFool  & 0.9955 & 0.9904 & 0.0096 \\
 \midrule
\multirow{3}{*}{\parbox{2cm}{Adversarially Trained Model}}
 &FGSM      & 0.9735 & 0.8977 & 0.1023 \\
&PGD       & 0.9735 & 0.9210 & 0.0790 \\
&DeepFool  & 0.9735 & 0.6649 & 0.3351 \\
\bottomrule
\end{tabular}%
\label{tab:performance-comparision-robustness}
}
\end{table}

% Reconstruction Error Figure
\begin{figure} [!h]
    \centering
    \subfloat[Clean vs. FGSM]{%
        \includegraphics[width=0.32\textwidth,keepaspectratio]{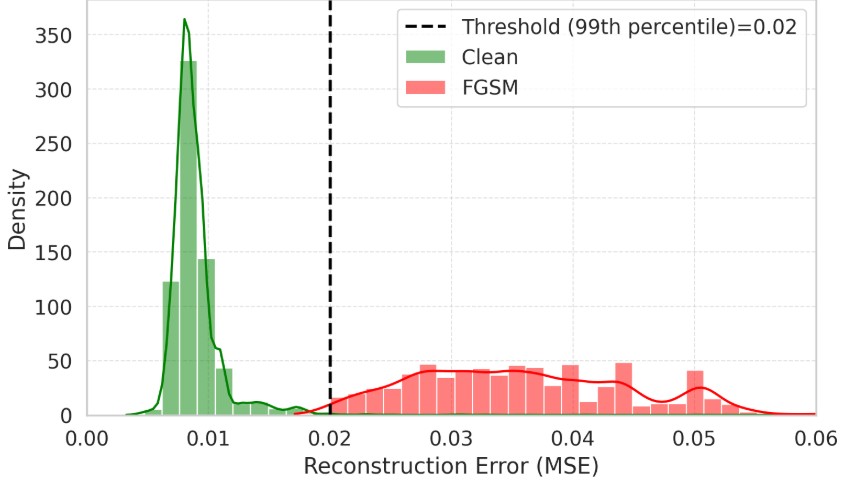}}
    \hfill
    \subfloat[Clean vs. PGD]{%
        \includegraphics[width=0.32\textwidth,keepaspectratio]{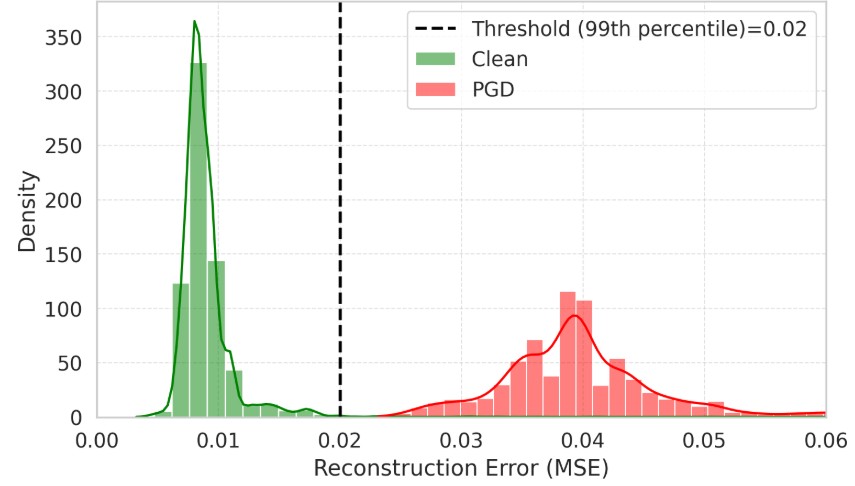}}
    \hfill
    \subfloat[Clean vs. DeepFool]{%
        \includegraphics[width=0.32\textwidth,keepaspectratio]{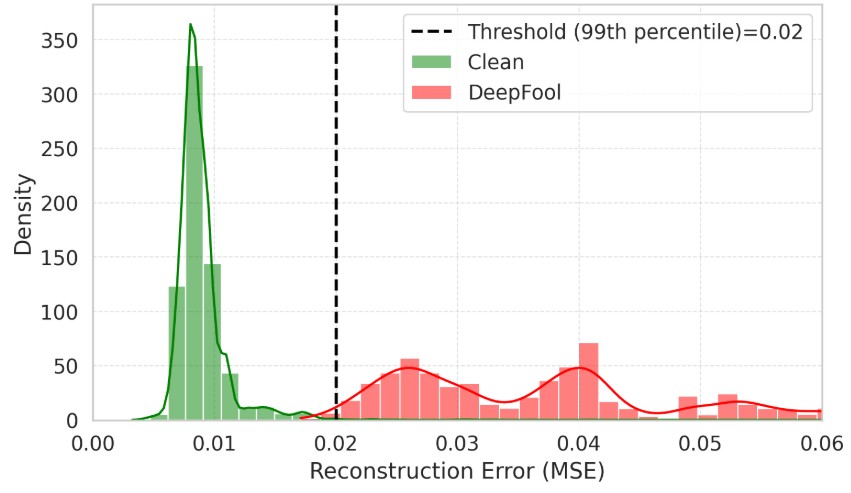}}
    \caption{\small Comparing Reconstruction Error Distribution of Clean and Adversarial Test Samples with Threshold $\tau=0.02$.}
    \label{fig:reconstruction-error}
\end{figure}
% Reconstruction error analysis
\subsubsection{Reconstruction Error Distribution \&  Analysis} \label{subsec:recon-error-analysis}
As our detector model is unsupervised and trained exclusively on clean data, we estimate a detection threshold \(\tau = 0.02\) based on $99^{th}$ percentile of reconstruction errors on the clean validation dataset to distinguish clean and adversarial inputs. We further analyze the reconstruction error distributions of both clean and adversarial samples against this threshold to assess the robustness and transparency of the decision boundary in our SHAP-based detection model.

Figs.~\ref{fig:reconstruction-error}  (a), (b), and (c) depict the distribution of reconstruction errors for clean versus FGSM, PGD, and DeepFool adversarial samples, respectively. These results show that the clean samples are tightly concentrated below \(\tau =0.02\), whereas adversarial samples exhibit significantly higher reconstruction errors, with a substantial portion extending well beyond the threshold \(\tau\). This clear separation demonstrates the model’s capability to distinguish adversarial inputs based on reconstruction deviation. In addition, this distribution plot provides direct insight into the model’s decision boundary, demonstrating how reconstruction error functions as a discriminative signal to classify inputs and support robust decision-making in adversarial settings.

%conclusion
\section{Conclusion \& Future Work} 	\label{sec:conclusion}
	In this study, we proposed a novel unsupervised adversarial detection model using a SHAP-based attribution fingerprinting to enhance the robustness of IoT intrusion detection against adversarial attacks. Our approach is suitable for IoT environments because it does not require retraining and relies solely on SHAP value computation, enabling fast and accurate adversarial detection in resource-constrained systems. We developed a complete data processing and attribution generation pipeline that includes adversarial sample creation and SHAP-based attribution computation for both clean and adversarial samples. Through attribution rank shift analysis, we revealed how different adversarial attacks alter feature importance, offering insight into attack-specific behavioral patterns. Experimental results on IoT benchmark dataset demonstrate that our model consistently outperforms a state-of-the-art adversarial defense method across FGSM, PGD, and DeepFool attacks, achieving superior performance in all classification, discrimination, and robustness metrics. By incorporating explainable AI, our proposed approach not only enhances adversarial detection but also improves model transparency and reliability in security-critical applications.

    We plan to conduct the following {\bf future work} to further improve  this SHAP-based attribution fingerprinting adversarial detection approach:
    \begin{itemize}
        \item Expanding from binary to multi-class intrusion detection across diverse IoT benchmark datasets for finer attack differentiation and better generalization.
        \item Incorporating adaptive online or continual learning to detect evolving, zero-day, or novel adversarial threats dynamically.
        \item Applying temporal and sequential modeling with attention or transformer architectures, while optimizing lightweight designs for real-time edge and fog deployment.
        
    \end{itemize}
    
\section{Acknowledgment} 
\small{This project is supported by collaborative research funding from the National Research Council of Canada’s Artificial Intelligence for Logistics Program.}
    
%Bibliography
	\bibliographystyle{IEEEtranSN}
	\small{
		\bibliography{references}
	}
\end{document}